\documentclass[5p,number,sort&compress,times,letterpaper]{elsarticle}

\usepackage[T1]{fontenc}
\usepackage{xcolor}
\usepackage{amsmath}
\usepackage{amssymb} 
\usepackage{bm}
\usepackage{newtxtext,newtxmath}  
\usepackage{siunitx}
\usepackage{url}
\usepackage{booktabs}
\usepackage{subfiles}
\graphicspath{%
  {./figs/}
}

\usepackage{stfloats, caption} 

\newcommand{\latin}[1]{\emph{#1}}
\newcommand{\angst}[0]{\r{a}ngstr\"om}
\newcommand{\subs}[1]{_{\mathrm{#1}}}
\newcommand{\supr}[1]{^{\mathrm{#1}}}

\journal{Journal of Magnetic Resonance}    

\begin{document}\begin{frontmatter}

\title{A theoretical and experimental assessment of adiabatic losses 
in force-gradient-detected magnetic resonance of nitroxide spin labels}

\author[1]{Michael C. Boucher\fnref{present1} \corref{cor1}}
\fntext[present1]{Present address: United States Naval Research Laboratory, Washington, District of Columbia 20375, USA}
\ead{mcb344@cornell.edu}
\author[1]{Peter Sun\fnref{present2}}
\fntext[present2]{Present address: Brookhaven National Laboratory, Upton, New York 11973, USA}
\author[1]{Eric W. Moore}
\author[1]{John A. Marohn \corref{cor1}}
\ead{jam99@cornell.edu}

\cortext[cor1]{Corresponding author}

\affiliation[1]{
  organization={Department of Chemistry and Chemical Biology, Cornell University}, 
  city={Ithaca},
  state={New York 14853},
  country={USA}
}

\begin{abstract}
We recently introduced a new theoretical description of Landau--Zener--St\"{u}ckelberg--Majorana (LZSM) transitions that accounts for both adiabatic and spin-dephasing losses during sweeps through resonance.
Here, we use this new description to assess signal loss due to cantilever tip motion in magnetic resonance force microscopy experiments on electron spins.
We derive equations for spin-induced cantilever frequency shifts that account for the time-dependent magnetization present during cantilever-synchronized periods of irradiation and relaxation.
We show that a frequency shift can be created by either a force- or force-gradient coupling mechanism, depending on the periodicity and timing of the microwave irradiation; the frequency shift decreases when the spin-lattice relaxation time becomes shorter than the cantilever oscillation period.
Equations were validated by comparing with the magnetization computed by numerically integrating the time-dependent Bloch equations.
Numerical simulations incorporating the new equations were compared to frequency-shift electron-spin signals collected as a function of magnetic field, tip-sample separation, microwave power, and microwave timing.
The simulations quantitatively describe the observed signals with independently calibrated parameters and no per-curve adjustment.
Finally, motivated by our new frequency-shift equations, we present a new experimental spin-excitation protocol that eliminates spurious signals arising from direct microwave excitation of the cantilever in a magnetic resonance force microscope experiment.
\end{abstract}


\end{frontmatter}

\section{Introduction}

Inspired by Rugar \latin{et al.}'s mechanical detection and imaging of electron spin resonance from a single atomic defect in quartz \cite{Rugar2004jul}, Moore and coworkers proposed and demonstrated the detection of magnetic resonance from biological nitroxide spin labels using a magnet-tipped cantilever \cite{Moore2009dec}.
Inspired by Moore, Nguyen and Marohn proposed and simulated an experiment to image individual nitroxides at sub-\angst{} resolution \cite{Nguyen2018feb}.
In the Moore and Nguyen experiments, changes in spin magnetization induced by resonant microwave irradiation are detected as a shift in the mechanical resonance frequency of an oscillating cantilever, and the per-spin sensitivity increases as the cantilever amplitude increases and the tip magnet diameter decreases.
Moore's experiments were carried out with a \SI{4}{\micro\meter}-diameter tip and reported quantitative agreement between measured and calculated signals \cite{Moore2009dec}.
In subsequent electron-spin experiments carried out with \SI{100}{\nano\meter}-diameter tips \cite{Hickman2010nov,Longenecker2011may}, calculated and measured signals differed by a factor of up to $400$ \cite{Hickman2010nov,Boucher2023jan}. 
Boucher \latin{et al.}\ improved this agreement $20$-fold by developing a protocol for applying a specially-prepared metal coating to the sample to decrease surface noise \cite{Boucher2023jan}.
The source of the remaining signal deficit is a topic of ongoing research.

One concern with the Moore--Nguyen electron-spin experiment is the presence of large time-dependent resonance offsets created by cantilever motion during irradiation.
In a recent paper, Boucher and coworkers modeled the breakdown of saturation and adiabatic rapid passage during cantilever motion by providing a new theoretical description of Landau--Zener--St\"{u}ckelberg--Majorana (LZSM) transitions that accounts for $T_2$ losses \cite{Boucher2023sep}.
They identified two unitless parameters that predict the ability of transverse irradiation, applied during a magnetic field sweep, to saturate or invert spin magnetization.
These parameters depend on the spin gyromagnetic ratio $\gamma$, 
rate of magnetic field change $d B_0 \big/ d t$, 
intensity of the oscillating transverse magnetic field $B_1$,
and spin dephasing time $T_2$.
They used physical and mathematical reasoning to derive new and simple analytical formulas for the post-sweep magnetization in the low-$B_1$ and high-$B_1$ limits when $T_2 \ll T_1$. 

In the course of incorporating these insights into numerical calculations of the frequency-shift signal in magnetic resonance force microscope experiments, we recently revisited numerical simulations of Moore's experiment \cite{Moore2009dec}, carried out with \SI{4}{\micro\meter}-diameter nickel tip. 
Performing the Ref.~\citenum{Moore2009dec} signal simulations more carefully, we found significant disagreement between theory and experiment.
Here, we show that this disagreement is resolved by including the swept-field breakdown of saturation in the signal calculation.

When the above approach was used to compute the signal in a small-tip electron-spin experiment, the computed signal was $15$-fold larger than the observed signal \cite{Boucher2023jan}. 
Prior studies of atomic defects in quartz raised concerns about signal loss from deleterious spin relaxation due to tip thermomagnetic fluctuations \cite{Hannay2000may,Stipe2001dec,Mozyrsky2003feb} and 
thermal current fluctuations \cite{Stipe2001dec,Kenny2021jun}.
In Ref.~\citenum{Boucher2023jan}, Boucher \latin{et al.}\ proposed that their nitroxide signal deficit could likewise be due to a reduction in electron-spin $T_1$ from large magnetic field fluctuations present near the magnetic tip.
The frequency shift signal employed so far to detect nitroxide spins has only been calculated in the limit that the electron's spin-lattice relaxation time $T_1$ is much longer than the cantilever period $T\subs{c} \sim \SI{200}{\micro\second}$, $T_1 \gg T\subs{c}$.
To assess Boucher's signal-deficit proposal, we need equations for the cantilever frequency shift applicable in the $T_1 \ll T\subs{c}$ limit.
Below, we derive equations for the spin-induced cantilever frequency shift valid in this limit.

\begin{figure*}
  \centering
  \includegraphics[width=5.5in]{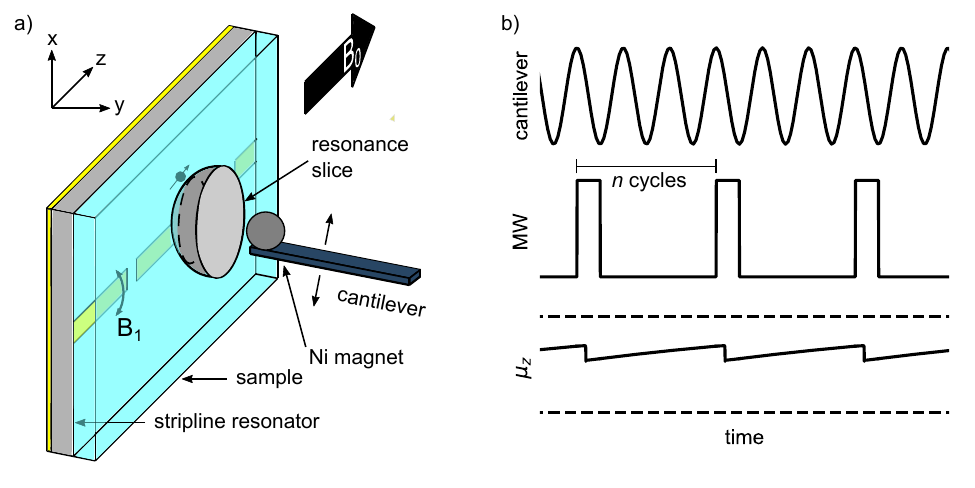}
  \caption{Magnetic resonance force microscope apparatus and microwave irradiation scheme.
    (a)  A \SI{40}{\milli\text{M}} sample of 4-amino-TEMPO in polystyrene was spin-coated onto a quartz substrate and placed over a microstripline resonator, the source of an oscillating magnetic field. 
    The cantilever was oscillated  in the $x$-direction, with a zero-to-peak amplitude of $x_\mathrm{pk} = \SI{164}{\nano\meter}$.
    Note that the $x$ direction is orthogonal to both the external field and the long axis of the cantilever \cite{Marohn1998dec}. 
    (b) Bursts of microwave (MW) irradiation were applied in synchrony with the cantilever oscillation.
    These bursts lasted for a half-cantilever-cycle duration and were typically applied every third cantilever cycle. 
    In (b), we depict the resulting $z$ component of spin magnetization $\mu_z$ versus time at a single sample location. 
  }
\label{fig:ESR_schematic}
\end{figure*}

The Ref.~\citenum{Moore2009dec} magnetic resonance force microscope (MRFM) experiment \cite{Sidles1991jun,Rugar1992dec} is sketched in Fig.~\ref{fig:ESR_schematic}.
Building on the Cantilever Enabled Readout of Magnetization Inversion Transients (CERMIT) experiment of Garner and coworkers \cite{Garner2004jun}, in Ref.~\citenum{Moore2009dec}, the authors detected electron spin magnetic resonance as a shift in the frequency of a magnetic-tipped cantilever.
The experimental setup --- including cantilever-sample orientation, static magnetic field, and oscillating microwave magnetic field --- is sketched in Fig.~\ref{fig:ESR_schematic}(a). 
To minimize field-induced changes in cantilever frequency and mechanical quality factor, the Ref.~\citenum{Moore2009dec} experiment was performed in the Springiness Preservation by Aligning Magnetization (SPAM) geometry \cite{Marohn1998dec}.
To evade surface frequency noise, the cantilever frequency was made time dependent by modulating the applied microwaves in an on--off sequence at a repetition rate of $f\subs{mod} = \SI{6.48}{\hertz}$.
In Fig.~\ref{fig:ESR_schematic}, we sketch the microwave-on period, during which microwave excitation was applied every $n$ cantilever cycles.

\section{Theory}

In the Fig.~\ref{fig:ESR_schematic} experiment, short irradiation periods are followed by resting periods during which the sample magnetization relaxes towards equilibrium.
We are interested in the time-dependent magnetization at \emph{steady-state}. 
In Sec.~\ref{sec:intermittent-weak} we derive, for constant resonance offset, formulas for the average magnetization that capture the equilibrium between evolution under irradiation and spin-lattice relaxation in the absence of irradiation.
In Sec.~\ref{sec:intermittent-strong} we adapt the results of Sec.~\ref{sec:intermittent-weak} to the case of time-dependent resonance offset.

\subsection{Constant radiation, constant resonance offset}
\label{sec:constant-weak}

With the microwaves on, assuming constant resonance offset, magnetization evolves toward a saturated steady-state value $M_z\supr{ss}$ at a rate $r$ given by the Bloch equations \cite{Boucher2023sep}:

\begin{equation}
M_z\supr{ss} 
 = 
 M_0 \frac{
   1 + \Omega^2}
 {1  + S^2 + \Omega^2},
 \label{eq:SS}
\end{equation}

\begin{equation}
r
=
\frac{1}{T_1} + \frac{S^2}{T_1\left(1 + \Omega^2\right)},
\end{equation}
with $S = \gamma B_1 \sqrt{T_1 T_2}$ and $\Omega = T_2 (\gamma B_0 - \omega) $ for the irradiation frequency $\omega$. The original simulations of Ref.~\citenum{Moore2009dec} used eq.~\ref{eq:SS} to approximate the resulting sample magnetization. 

\subsection{Intermittent radiation, constant resonance offset}
\label{sec:intermittent-weak}

\begin{table}
\begin{center}
\renewcommand{\arraystretch}{1.50}
\begin{tabular}{ccc}\toprule
MW
 & duration 
 & magnetization \\ \hline
on 
 & $\tau\subs{on}$ 
 & $M_z\supr{on}(t) 
   =(M_z(0^{-}) - M_z\supr{ss}) 
   \, e^{-r t} + M_z\supr{ss}$ \\
off 
  & $\tau\subs{off}$ 
  & $M_z\supr{off}(t) 
    = (M_z(0^{+}) - M_0) 
    \, e^{-r_1 t} + M_0$ \\ \bottomrule
\end{tabular}
\end{center}
\caption[Magnetization evolution during intermittent irradiation]{
  Magnetization evolution during microwave-on and microwave-off time periods. 
  In the magnetization column, for simplicity, we have defined time $t$ relative to the start of \emph{each} period.}
\label{table:irradiation}
\end{table}

With the microwaves off, magnetization evolves towards an equilibrium value $M_0$ at a rate $r_1 = 1/T_1$. 
The irradiation scheme and resulting magnetization dynamics are summarized in Table~\ref{table:irradiation}.
In the table, $M_z(0^{-})$ and $M_z(0^{+})$ are the magnetization just prior to the start of the microwave-on and microwave-off period, respectively.

At \emph{steady state} we require the magnetization to be continuous:
\begin{subequations}
\begin{align}
M_z\supr{on}(\tau\subs{on}) & = M_z(0^{+}) \text{ and}\\
M_z\supr{off}(\tau\subs{off}) & = M_z(0^{-}).
\end{align}
\end{subequations}
These continuity equations can be used to eliminate the unknown variables $M_z(0^{-})$ and $M_z(0^{+})$ from the Table~\ref{table:irradiation} equations as follows.
Substituting
\begin{subequations}
\begin{align}
(M_z(0^{-}) - M_z\supr{ss}) 
 \, E\subs{on} + M_z\supr{ss} & = M_z(0^{+})
 \label{eq:Mz-contin-1} \text{ and}\\
(M_z(0^{+}) - M_0) 
 \, E\subs{off} + M_0 & = M_z(0^{-}),
 \label{eq:Mz-contin-2}
\end{align}
\end{subequations}
with
\begin{subequations}
\begin{align}
E\subs{on} \equiv & e^{-r \, \tau\subs{on}}
\text{ and }\\
E\subs{off} \equiv & e^{-r_1 \, \tau\subs{off}},
\end{align} 
\end{subequations}
and then inserting the eq.~\ref{eq:Mz-contin-1} expression for $M_z(0^{+})$ into eq.~\ref{eq:Mz-contin-2} and solving for $M_z(0^{-})$ gives
\begin{equation}
M_z(0^{-}) 
 = 
 \frac{
   E\subs{off} (1 - E\subs{on}) M_z\supr{ss}
   + (1 - E\subs{off}) M_0
 }
 {1 - E\subs{on} \, E\subs{off}}.
 \label{eq:Mz0-}
\end{equation}
Likewise, plugging the eq.~\ref{eq:Mz-contin-2} expression for $M_z(0^{-})$ into eq.~\ref{eq:Mz-contin-1} yields
\begin{equation}
M_z(0^{+}) 
 = 
 \frac{
   (1 - E\subs{on}) M_z\supr{ss}
   + E\subs{on} (1 - E\subs{off}) M_0
 }
 {1 - E\subs{on} \, E\subs{off}}.
 \label{eq:Mz0+}
\end{equation}

The measured signal is proportional to the difference between the microwave-modulated magnetization and the equilibrium magnetization.
During the microwave-on period, $\Delta M_z\supr{on}(t) = M_z\supr{on}(t) - M_0$.
Substituting eq.~\ref{eq:Mz0-} for $M_z(0^{-})$, we obtain
\begin{equation}
\Delta M_z\supr{on}(t)
 = (M_z\supr{ss} - M_0)
 - \frac{1 - E\subs{off}}
    {1 - E\subs{on} E\subs{off}}
    (M_z\supr{ss} - M_0) \, e^{-r t}.
    \label{eq:DMz-on} 
\end{equation}
During the microwave-off period, $\Delta M_z\supr{off}(t) = M_z\supr{off}(t) - M_0$.
Substituting eq.~\ref{eq:Mz0+} for $M_z(0^{+})$, we obtain
\begin{equation}
\Delta M_z\supr{off}(t)
 = \frac{1 - E\subs{on}}
    {1 - E\subs{on} E\subs{off}}
    (M_z\supr{ss} - M_0) \, e^{-r_1 t}.
    \label{eq:DMz-off} 
\end{equation}
These answers are physically reasonable.
When the irradiation--off period is infinitely long, $E\subs{off} \rightarrow 0$ and $\Delta M_z\supr{on}(t) = (M_z\supr{ss} - M_0)(1 - e^{-r t})$; the signal is zero initially and rises to $M_z\supr{ss} - M_0$ as $t \rightarrow \infty$.
When the irradiation-on period is infinitely long, $E\subs{on} \rightarrow 0$ and $\Delta M_z\supr{off}(t) = (M_z\supr{ss} - M_0) \, e^{-r_1 t}$; a large initial signal is created that decays exponentially to zero as $t \rightarrow \infty$.

The time-averaged signal in our MRFM experiment is proportional to
\begin{equation} 
\langle \Delta M_z \rangle
 = \frac{1}{\tau\subs{on}+\tau\subs{off}}
 \left(
 \int_{0}^{\tau\subs{on}}
   \Delta M_z\supr{on}(t) \, dt \right. \\
 \left.
 +
 \int_{0}^{\tau\subs{off}}
   \Delta M_z\supr{off}(t) \, dt 
 \right).
\end{equation} 
Inserting eqs.~\ref{eq:DMz-on} and \ref{eq:DMz-off} into the above equation and carrying out the integrals yields the following result for the average magnetization:
\begin{equation}
\langle \Delta M_z \rangle 
 = \frac{M_z\supr{ss} - M_0}
    {\tau\subs{on}+\tau\subs{off}} 
 \left(
 \frac{(1 - E\subs{on})(1 - E\subs{off})}
      {1 - E\subs{on} \, E\subs{off}}
 \frac{r - r_1}{r r_1} 
 + \tau\subs{on}
 \right).
 \label{eq:DeltaMz-weak-B1}
\end{equation}

\subsection{Intermittent radiation, time-dependent resonance offset}
\label{sec:intermittent-strong}

For experiments with a time-dependent resonance offset, the steady-state value $M_z\supr{ss}$, and relaxation rate $r$ are no longer constants, making it much more difficult to solve the continuity equations. 
However, for most measurements of interest, spins remain near resonance for a period much shorter than the spin-lattice relaxation time, $\tau\subs{on} \ll T_1$, and much shorter than the time between microwave pulses, $\tau\subs{on} \ll \tau\subs{off}$. 

For sufficiently large  $B_1$,
\begin{equation}
M_z(0^{+}) \approx f\subs{on} \, M_z(0^{-}),
\label{eq:Mz-contin-3}
\end{equation}
where $f\subs{on}$ is $-1$ for adiabatic rapid passage, $0$ for a perfectly saturating sweep, and is $1$ when saturation is lost due to moving-tip effects.
In eq.~\ref{eq:Mz-contin-3} we assume that spins equilibrate with the microwaves instantaneously, valid when $\tau\subs{on} \ll \tau\subs{off}$.
We further assume that $M_z\supr{sat}$ is zero, valid when $B_1 \gg B\subs{sat}$.
By setting $E\subs{on} \rightarrow f\subs{on}$, $M_z\supr{ss} \rightarrow 0$, $\tau\subs{on} \rightarrow 0$, and assuming $r \gg r_1$ in eq.~\ref{eq:DeltaMz-weak-B1}, we obtain the following approximation for the average magnetization:
\begin{equation}
\langle \Delta M_z \rangle 
 \approx -\frac{M_0}{r_1 \tau\subs{off}}
 \frac{(1 - f\subs{on})(1 - E\subs{off})}
      {1 - f\subs{on} \, E\subs{off}}, 
 \label{eq:DeltaMz-strong-B1}
\end{equation}
with $f\subs{on} = m_z\supr{final} \big/ m_z\supr{initial}$ the ratio of the magnetization after the sweep to the magnetization before the sweep.
Boucher and coworkers found that this magnetization ratio depended on two unitless parameters
\begin{subequations}
\begin{align}
\alpha_1 & \equiv \frac{1}{\pi \gamma B_1^2} \left|\frac{d \Delta B_0}{d t}\right| \text{ and}\\
\alpha_2 & \equiv \frac{T_2}{\pi B_1} \left|\frac{d \Delta B_0}{d t}\right|.
\end{align}\label{eq:alphas}
\end{subequations} 
In Ref.~\citenum{Boucher2023sep}, analytical approximations for the magnetization ratio were computed in two limits.
The limits depend on the electron-spin resonance parameters, given for the nitroxide 4-amino-TEMPO in Table~\ref{table:parameters}.
When $B_1 \ll B_1\supr{crit}$, the transverse field is too small to achieve a Rabi oscillation.  
In this limit
\begin{equation}
m_z\supr{final} 
  \approx e^{-1 \big/ \alpha_1}\, m_z\supr{initial}.
\label{eq:lowB1}
\end{equation}
On the other hand, when $B_1 \gg B_1\supr{crit}$,
\begin{equation}
m\supr{final}_z 
  \approx e^{-1 \big/ \alpha_2} \left( -1 + 2 \, e^{-1 \big/ 2 \alpha_1} \right) \, m_z\supr{initial}.
  \label{eq:L-Z-guess}
\end{equation}
In the simulations described in this work we use eq.~\ref{eq:DeltaMz-strong-B1}, with $f\subs{on}$ given by eq.~\ref{eq:lowB1}. In Figs.~\ref{fig:Vary_power_fit_cP}, ~\ref{fig:Vary_power_vs_field}, and ~\ref{fig:Vary_interpulse_spacing}(b) simulations are carried out where $B_1$ exceeds $B_1\supr{crit}$. However, as shown in Figs.~S1 and S2, the ``Low-$B_1$ limit'' eq.~\ref{eq:lowB1} produces a result very close to the full Bloch equations, and a better result than eq.~\ref{eq:L-Z-guess}.

In the case where the cantilever motion is small compared to the diameter of the magnetic tip, the contribution of a single spin at location $\bm{r}$ can be calculated as:
\begin{equation}
  \Delta f\subs{spin}  
    = - \frac{f_c \langle \Delta M_z \rangle}{2 k_c}  
       \frac{\partial^2  B\supr{tip}_z}{\partial x^2}(\bm{r}).
\end{equation}
 
\begin{table}
\centering
\begin{tabular}{l@{\hspace*{1.5em}}l@{\hspace*{1.5em}}l} \toprule
variable & definition & numerical value \\ \midrule
$\gamma$ & & $ 2\pi \times \SI{28.0}{\giga\hertz\per\tesla}$ \\
$T_2$ & $1/r_2$ & $\SI{0.45}{\micro\second}$ \\
$T_1$ & $1/r_1$ & $\SI{1.3}{\milli\second}$ \\
$B\subs{sat}$ & $1/(\gamma \sqrt{T_1 T_2})$ & $\SI{0.24}{\micro\tesla}$ \\
$B_1\supr{crit}$ & $(r_2 - r_1)/(2 \gamma )$ & $\SI{6.3}{\micro\tesla}$  \\ \bottomrule
\end{tabular}
\caption{Electron spin resonance parameters for 4-amino-TEMPO. $T_1$ and $T_2$  come from Ref.~\citenum{Moore2009dec}. They were obtained via inductive measurements at 4.2 K and 17.7 GHz on a 40 mM doped polystyrene film.}
\label{table:parameters}
\end{table}

\subsection{Intermittent radiation, short $T_1$ limit}
\label{sec:intermittent-strong-short}

Above we used the cycle-averaged change in magnetization $\langle \Delta M_z \rangle$ to compute the spin-induced cantilever frequency shift $\Delta f\subs{spin}$.
Using the cycle-averaged magnetization in the frequency-shift calculation is an approximation, valid in the $T_1 \gg T\subs{c}$ limit, with $T\subs{c}$ the cantilever period.
Here we relax this assumption, deriving an equation for $\Delta f\subs{spin}$ valid even in the $T_1 \ll T\subs{c}$ limit.

The starting point for this derivation is Eq.~1 in Ref.~\citenum{Lee2012apra}, which tells us how a time-dependent tip force produces a cantilever frequency shift: 
\begin{equation}
  \Delta f_c = - \frac{f_c}{k q\subs{pk}^2} \langle \bm{F}\subs{tip} \cdot \bm{q} \rangle_T.
  \label{eq:Lee}
\end{equation}
Here $f_c$ is the cantilever frequency, $k_c$ is the spring constant of the cantilever, $q\subs{pk}$ is the amplitude of the cantilever from zero to peak, $\bm{q} = q(t) \, \hat{\bm{x}}$ is the displacement of the tip with
\begin{equation}
  q(t) = q\subs{pk} \sin{(2 \pi f_c t)},
\end{equation}
and $\bm{F}\subs{tip}$ is the time-dependent force acting on the tip.
In eq.~\ref{eq:Lee}, the bracket $\langle \rangle_T$ indicates a temporal average taken over one period of oscillation $T$ of the force.
In the calculations below, $T$ will be a multiple of the cantilever period $T\subs{c}$.

Consider a single spin at location $\bm{r}$.
As the cantilever oscillates, $\bm{r} \rightarrow \bm{r} - q(t) \, \hat{\bm{x}}$, and 
$\bm{F}\subs{tip} \cdot \bm{q} = F_x(\bm{r} - q(t) \, \hat{\bm{x}})q(t)$, where $F_x(\bm{r}) = -M_z(t) \: \partial B\supr{tip}_z \big/ \partial x$ is the force on the tip.
Expanding $\bm{F}\subs{tip} \cdot \bm{q}$ to first order in $q(t)$, valid when $|q(t)| \ll r$, gives
\begin{equation}
\bm{F}\subs{tip} \cdot \bm{q} 
  \approx
  -M_z(t) \left( G_1 - q(t) G_2\right) q(t)
  \label{eq:Lee-linearized}
\end{equation}
with
\begin{subequations}
\begin{align}
G_1 & \equiv \frac{\partial  B\supr{tip}_z}{\partial x}(\bm{r}) \text{ and}\\
G_2 & \equiv \frac{\partial^2  B\supr{tip}_z}{\partial x^2}(\bm{r}), 
\end{align}
\end{subequations}
the first and second derivatives, respectively, of the tip field; 
the derivatives are taken in the direction of cantilever motion.
The $G_1$ term leads to signal in the OScillating Cantilever-driven Adiabatic Reversals (OSCAR) experiment \cite{Stipe2001dec,Mamin2003nov,Rugar2004jul}, 
whereas the $G_2$ term leads to signal in the CERMIT experiment \cite{Garner2004jun,Moore2009dec,Lee2012apra,deWit2019feb}.

Let us use the above equations to compute the baseline frequency shift $\Delta f\subs{rest}$ arising from a spin at thermal equilibrium interacting with a magnetic-tipped cantilever.  With $M_z(t) = M_0$, 
\begin{equation}
\Delta f_{\subs{\mathrm{rest}}} 
= \frac{f_c M_0}{k_c q_{\subs{\mathrm{pk}}}^2 T_{\subs{\mathrm{c}}}}
\int_0^{T_{\subs{\mathrm{c}}}} \! \! dt  
\: \left( G_1 - q_{\subs{\mathrm{pk}}}  \sin(\frac{2 \pi t}{T_{\subs{\mathrm{c}}}}) \, G_2 \right)
q_{\subs{\mathrm{pk}}} \sin(\frac{2 \pi t}{T_{\subs{\mathrm{c}}}}),
\end{equation}
where $T\subs{c}$ is the cantilever period.  This integral evaluates to
\begin{equation}
  \Delta f\subs{rest}  
    = - \frac{f_c M_0}{2 k_c}  
       \frac{\partial^2  B\supr{tip}_z}{\partial x^2}(\bm{r}).
\end{equation}

\begin{figure}
  \centering
  \includegraphics[width=2.75in]{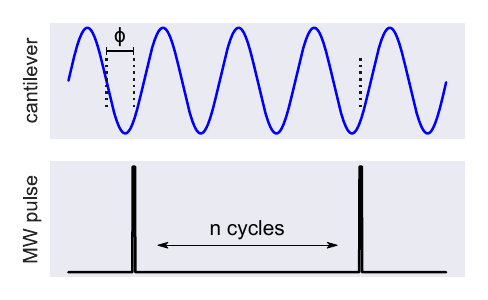}
  \caption{Time-delayed intermittent microwave bursts applied every $n$ cantilever cycles.
    Cantilever position (upper) and microwave intensity (lower) are plotted versus time, with 
    the microwave phase delay indicated as $\phi$.
  }
  \label{fig:arb_phase_pulse_sequence}
\end{figure}

Now consider the change in cantilever frequency $\Delta f\subs{MW}$ induced by bursts of microwave irradiation delivered every $n$ cantilever cycles.
The irradiation timing diagram is shown in Fig.~\ref{fig:arb_phase_pulse_sequence}.
The periodic microwave bursts, while synchronized with the cantilever oscillation, are time-delayed by a fraction of a cantilever period.
The time delay is equivalent to a phase shift $\phi$. 
The magnetization following a microwave burst at $t = 0$ is given by
\begin{equation}
M_z(t) = M_0 + (M_{z}\supr{ss} - M_0)e^{-r_1 t},
\label{eq:Mzt-arb-phase}
\end{equation}
with $r_1 = 1/T_1$ the spin-lattice relaxation rate.
As in Sec.~\ref{sec:intermittent-strong}, in eq.~\ref{eq:Mzt-arb-phase} we assumed that the magnetization reaches steady state instantaneously upon application of microwaves.
The change in cantilever frequency is computed as
\begin{multline}
\Delta f\subs{MW}
 = \frac{f_c}{k_c q^2\subs{pk} n T\subs{c}}
 \int_0^{n T\subs{c}} dt \: M_z(t) \: \times \\
 \left( 
   q\subs{pk} G_1 \sin{(\frac{2 \pi t}{T\subs{c}} + \phi)}
   - q\subs{pk}^2 G_2 \sin^2{(\frac{2 \pi t}{T\subs{c}} + \phi)}
 \right). 
\end{multline}
After evaluating the above integral, the signal is computed as
\begin{equation}
\Delta f\subs{spin} = \Delta f\subs{MW} - \Delta f\subs{rest}.
\end{equation}
For the Fig.~\ref{fig:arb_phase_pulse_sequence} irradiation scheme we find
\begin{multline}
\Delta f\subs{spin} = 
-\frac{f_c}{k_c}
  (M_z\supr{ss} - M_0)
   (1 - e^{-r_1 T}) \\
   \times \left\{
   -\frac{G_1}{q_\mathrm{pk}} 
   \frac{2 \pi n \cos(\phi) + r_1 T \sin(\phi)}
          {4 \pi^2 n^2  + (r_1 T)^2} \right. \hspace{0.75in} \\
    \left. + 
    \frac{G_2}{2}
    \left(
      \frac{1}{r_1 T} 
       - \frac{r_1 T \cos(2 \phi) - 4 \pi n \sin(2 \phi)}
                 {16 \pi^2 n^2 + (r_1 T)^2}
    \right) \right\},
\label{eq:Df_vary_phase}
\end{multline}
with $T = n T\subs{c}$ the irradiation repetition period.

\begin{figure}
  \centering
  \includegraphics[width=2.75in]{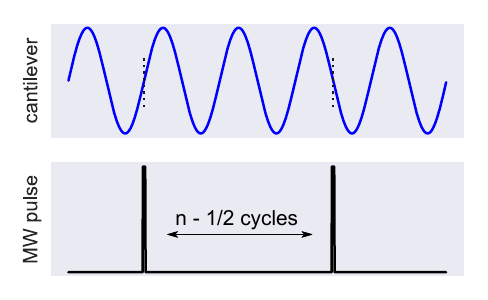}
  \caption{Intermittent microwave bursts applied every $(n - \frac{1}{2})$ cantilever cycles, with $n \geq 1$.
    Cantilever position (upper) and microwave intensity (lower) are plotted versus time.
  }
  \label{fig:n_half_delay_pulse}
\end{figure}

Consider instead the irradiation timing diagram shown in Fig.~\ref{fig:n_half_delay_pulse}.
Here we apply microwave bursts at alternating cantilever zero crossings (i.e.\ the midpoint of the cantilever oscillation). This restricts time between microwave bursts to $(n - \frac{1}{2}) T\subs{c}$ with $n \geq 1$. 
After two bursts, the magnetization is again in phase with the cantilever, so the irradiation repetition period is $T = (2 n - 1) T\subs{c}$.
For the Fig.~\ref{fig:n_half_delay_pulse} irradiation scheme we find (see Supporting Information for derivation)
\begin{multline}
\Delta f\subs{spin} = 
-\frac{f_c G_2}{k_c r_1 T}
  (M_z\supr{ss} - M_0)
   (1 - e^{-r_1 T/2}) \\
   \times \frac{16 \pi^2 (2n - 1)^2}
          {16 \pi^2 (2n - 1)^2 + (r_1 T)^2}.
\label{eq:Df_half_delay}
\end{multline}

\section{Methods}

\subsection{Experiment details}

The measurements of Figs.~\ref{fig:Vary_tip_sample_Moore}, \ref{fig:Vary_power_fit_cP}, \ref{fig:Vary_power_vs_field}, and \ref{fig:Vary_interpulse_spacing} were performed using the apparatus and protocols described in Ref.~\citenum{Moore2009dec}. For these experiments the sample was a \SI{200}{\nano\meter} thick polystyrene (Polymer Source, P4179B-PS, $M_n = 200\times10^3$ and $M_w/M_n = 1.4$) film doped to \SI{40}{\milli\text{M}} with 4-amino-TEMPO (Aldrich, 163945). 
The polymer was spin-coated (2000 rpm, 30 s) onto a quartz substrate (NOVA Electronics) using deuterated toluene as a solvent. 
A \SI{20}{\nano\meter} thick layer of gold was e-beam evaporated onto the sample at a rate of \SI{0.2}{\nano\meter\per\second}.

A cantilever with dimensions $\SI{200}{\micro\meter} \times \SI{4}{\micro\meter} \times \SI{340}{\nano\meter}$ was fabricated as described in Jenkins \emph{et al.} \cite{Jenkins2004may}. 
A nickel sphere (Novamet, CNS-10) was glued to the end of the cantilever using Miller-Stevenson 907 epoxy. 
A scanning electron micrograph of the nickel tip was taken, and the tip was measured to have a radius of \SI{2.1}{\micro\meter}. 
The tip position was observed using a temperature-tuned \SI{1310}{\nano\meter} fiber-optic interferometer \cite{Rugar1989dec,Bruland1999sep}. 
The interferometer was aimed at an octagonal pad \SI{75}{\micro\meter} from the cantilever end.
The cantilever spring constant was estimated by observing the cantilever thermomechanical fluctuations and calculated using the equipartition theorem, $k_c = k_B T/\langle x^2_\mathrm{th} \rangle$. 
The calculated value was divided by 2.02 to account for the additional deflection of the cantilever tip relative to the cantilever pad \cite{Hickman2010feb}. 
The Brownian motion-derived spring constant measured in this way, $k_c = \SI{0.8}{\milli\newton\per\meter}$, was used in all simulations, except  Fig.~\ref{fig:Vary_tip_sample_Moore}(a), where the spring constant was increased to achieve improved agreement between theory and experiment.

An Anritsu-Wiltron source (model 6814B) was used as a microwave source. The microwave output was modulated by an American Microwave Corporation switch (model SWN-218-2DT) and amplified by a Narda microwave amplifier (model DBP-0618N830).
The microwave source power was typically set to \SI{-10}{\text{dBm}}, gained 28.5 dB from the Narda amplifier and lost 8.5 dB in cabling between the amplifier and microstripline resonator. 
The microwave frequency was \SI{18.1}{\giga\hertz}. 
The microstripline resonator used was fabricated on a PCB. The copper resonator was \SI{5.6}{\milli\meter} long, \SI{1.4}{\milli\meter} wide and \SI{36}{\micro\meter} thick. 
It had a \SI{20}{} to \SI{40}{\micro\meter} coupling gap, and a \SI{0.7}{\milli\meter} thick epoxy-glass dielectric. 
Outside the microscope, the resonator quality factor was measured to be $\sim 800$ using an RF power detector and a directional coupler (Anritsu-Wiltron 75KA50, and Krytar 102020020).

Magnetic resonance force microscopy was performed in the SPAM (Springiness Preservation by Aligning Magnetization) geometry \cite{Marohn1998dec,Garner2004jun}, Fig.~\ref{fig:ESR_schematic}, to avoid magnetic field losses of cantilever quality factor and force sensitivity. 
That is, the external field was pointed in the $z$-direction, the length of the cantilever and the tip-sample separation were in the $y$-direction, and the cantilever motion was in the $x$-direction. 
This geometry produces a signal both upfield and downfield of the Larmor resonance condition. The cantilever tip was positioned by an Attocube ANPx51/HV/LT. 
A \SI{1310}{\nano\meter} fiber interferometer was used to calibrate the positioner, and a set DC voltage was used to vary tip-sample separation. The MRFM probe was cooled by dunking the vacuum can in liquid helium, and the probe temperature was measured using a thermistor to be \SI{8}{\kelvin}. The simulation temperature was scaled to \SI{11}{\kelvin} to match the signal amplitude. The thermistor was not connected directly to the sample and the discrepancy might represent imperfect heat-sinking.

The cantilever was self-oscillated using a home-built positive feedback circuit. 
The positive feedback circuit phase shifted the cantilever signal by $-90$ degrees and converted it to a square wave with an amplitude set by a DC input. 
The resulting oscillation was then filtered with a second bandpass filter centered at the cantilever frequency. 
The microwave pulse trigger was produced by digitally dividing down the square wave signal to produce a signal at the modulation frequency, $f_\mathrm{mod}$, and a signal at the pulse repetition rate. 
These two signals were AND-gated together and used to trigger a Berkeley Nucleonics 565 pulse generator. 
Typically, the pulse repetition rate was chosen such that a microwave pulse occurred every third cantilever cycle, and $f_\mathrm{mod}$ was typically between $5$ and \SI{100}{\hertz}.
Figure~\ref{fig:Vary_interpulse_spacing} shows a signal in which the spacing between the microwave pulses was varied. 

The experimental signal was observed to be shifted $-\SI{5}{\milli\tesla}$ relative to the simulation. 
The most likely explanation is that the effective coil constant for the external field magnet was $<1\%$ larger than reported by the manufacturer at the location of the sample and cantilever. 
The simulated data were shifted $-\SI{5}{\milli\tesla}$ in Figures~\ref{fig:Vary_tip_sample_Moore}, \ref{fig:Vary_power_vs_field}, and \ref{fig:Vary_interpulse_spacing} to adjust for the error. 

The measurements of Fig.~\ref{fig:One_vs_both_crossings} and \ref{fig:Pulse_width_and_power_vs_lockin_signal} were performed using the probe described in Ref.~\citenum{Boucher2023jan}. Specifically, MRFM was performed in the hangdown geometry, with the cantilever normal to the sample and parallel to the external field. The sample was a \SI{300}{\nano\meter} thick polystyrene thin film doped to \SI{40}{\milli\text{M}} with 4-amino-TEMPO, prepared on a coplanar waveguide using the described lamination procedure. The cantilever was a \SI{200}{\micro\meter} silicon cantilever with a $\SI{100}{\nano\meter}\times\SI{70}{\nano\meter}\times\SI{1500}{\nano\meter}$ cobalt nanomagnet (cantilever A from Table 1 of reference~\citenum{Boucher2023jan}). 

\subsection{Simulation details}

Simulations of the MRFM signal were performed using the \texttt{mrfmsim} Python package \cite{Sun2023jul,Sun2026jun,Sun2024jan}; details follow.  
The average magnetization was calculated according to eq.~\ref{eq:DeltaMz-strong-B1}. 
The frequency shift signal was then calculated according to eq.~20 in Ref.~\citenum{Lee2012apra} by numerically integrating over a simulation sample grid. 
The magnetic tip field in the z-direction, $B^{\mathrm{tip}}_\mathrm{z}$, and its first derivative in the direction of cantilever motion, $B^{\mathrm{tip}}_{\mathrm{zx}}$, were calculated by assuming a uniformly polarized sphere. 
The radius $r$ and saturation magnetization $B_\mathrm{mag}$ were obtained from a fit of the measured signal \emph{vs}.\ tip-sample separation (Fig.~\ref{fig:Vary_tip_sample_Moore} inset). The tip velocity was assumed to be the maximum velocity during the cantilever cycle, $v_\mathrm{tip} = 2 \pi f_c x_\mathrm{pk}$, calculated using the cantilever zero-to-peak amplitude $x_\mathrm{pk}$ and frequency $f_c$. While this assumption is an approximation, most active spins pass through resonance when the cantilever is moving at close to its maximum velocity.

The simulation spin-lattice relaxation time used was $T_1 = \SI{1.3}{\milli\second}$, the average value of $T_1$ obtained on a similar sample in reference \citenum{Moore2009dec}. 
For most experiments, microwaves were pulsed every third cantilever cycle, so the value of $\Delta t$ used was \SI{544}{\micro\second}, three cantilever periods. 
The assumption being made was that the period where the spins were saturating was a negligible fraction of the time between microwave pulses, which was true for most sample spins. 
The simulations of Figure~\ref{fig:Vary_interpulse_spacing} were handled similarly, with $\Delta t$ set to be the appropriate multiple of the cantilever period. 

The sample was treated as a rectangular prism with an $(x,y,z)$ extent of ($\SI{20000}{\nano\meter}$, $\SI{200}{\nano\meter}$, $\SI{30000}{\nano\meter}$) and voxel dimensions ($\SI{20}{\nano\meter}$, $\SI{20}{\nano\meter}$, $\SI{30}{\nano\meter}$). 
In most experiments, the estimated microwave power at the resonator was \SI{10}{\text{dBm}}, with the coil constant fitted in Fig.~\ref{fig:Vary_power_fit_cP}, this would be expected to produce $B_1 = \SI{2.9}{\micro\tesla}$. 
This $B_1$ was used in all simulations except where otherwise noted. The simulation temperature was scaled to \SI{11}{\kelvin} to match the signal amplitude. The thermistor was not connected directly to the sample and the discrepancy between the \SI{8}{\kelvin} thermistor reading and \SI{11}{\kelvin} at the sample might represent imperfect heat-sinking.

The numerical simulations of the full Bloch equations of Figure~\ref{fig:Linearity_plot} were carried out using the \texttt{odeint} method of the \texttt{SciPy} Python package with a \SI{10}{\pico\second} time step.

\section{Results and discussion}

\begin{figure*}
\centering
    \includegraphics[width=6.0in]{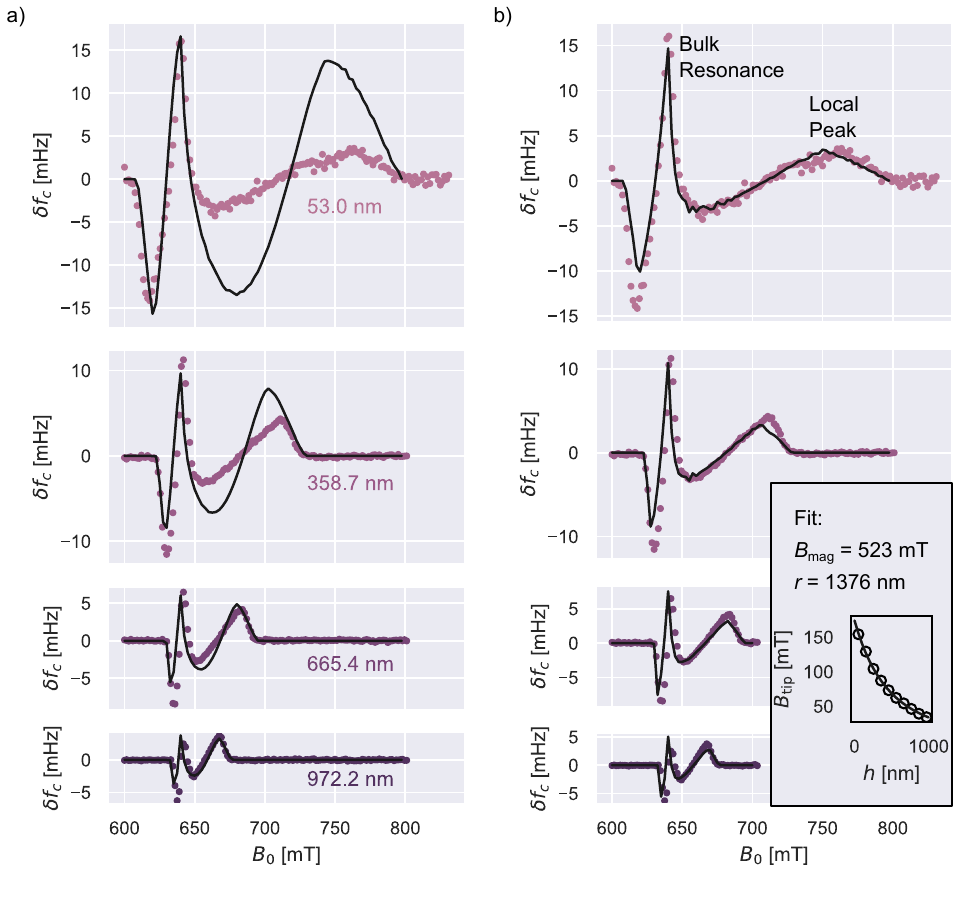}
    \vspace{-2em}
    \caption{Electron spin resonance signal \latin{vs}.\ external field $B_0$ at four tip-sample separations (colored circles), and signal simulations calculated using two different methods (black lines). 
    (a) Saturation calculated using the steady-state Bloch equations \cite{Moore2009dec} with $B_1 = \SI{2.9}{\micro\tesla}$ and $k_c = \SI{3.9}{\milli\newton\per\meter}$. For this method, the simulation cantilever spring constant was increased to attempt to better match the data. 
    (b) Saturation calculated according to eq.~\ref{eq:DeltaMz-strong-B1} with $B_1 = \SI{2.9}{\micro\tesla}$, $k_c = \SI{0.8}{\milli\newton\per\meter}$, and velocity assumed to be the maximum velocity during the \SI{164}{\nano\meter} amplitude motion of the cantilever. The observed bulk resonance at \SI{640}{\milli\tesla} and local peak are labeled.
    Inset: Tip field inferred from the data at each height (circles) and the best fit to eq.~\ref{eq:Btip-fit}. Inset residuals are plotted in Fig. S3. 
}
\label{fig:Vary_tip_sample_Moore}
\end{figure*}

We begin by establishing the magnetic properties of the cantilever tip.
In Fig.~\ref{fig:Vary_tip_sample_Moore}, we show the spin signal (colored dots) acquired at a number of different tip-sample separations. 
The signal shows an upfield ``local peak'' arising from spins in resonance directly below the tip and a downfield ``bulk peak'' arising from spins in resonance far away from the tip.
The maximum tip field $B_\mathrm{tip}$ at each tip-sample separation $h$ was calculated as the difference between the high-field edge of the ``local peak'' and the \SI{640}{\milli\tesla} observed resonant field of bulk spins. 
The Fig.~\ref{fig:Vary_tip_sample_Moore}(b) inset shows a fit of $B\subs{tip}$ \latin{vs.}\ $h$ data to
\begin{equation}
B_\mathrm{tip} = \frac{B_\mathrm{mag}}{3}\left(\frac{r}{r + h}\right)^3,
\label{eq:Btip-fit}
\end{equation}
the magnetic field expected at the equator of a uniformly magnetized sphere.
The best fit saturation magnetization was $B_\mathrm{mag} = \SI{523}{\milli\tesla}$ and best-fit radius was $r = \SI{1376}{\nano\meter}$. The best-fit value of the radius differs noticeably from the scanning electron micrograph. We attribute this discrepancy to the irregular shape of the particle, and note that the micrograph only shows the particle from one side.
These best-fit values were used in all subsequent simulations. 

Figure~\ref{fig:Vary_tip_sample_Moore} compares two different methods of calculating sample saturation. 
To compute the signal in Fig.~\ref{fig:Vary_tip_sample_Moore}(a), at each sample grid point, we determined the minimum resonance offset during the microwave pulse and used that offset, along with the steady-state Bloch equations result, to compute the magnetization.
To compute the signal in Fig.~\ref{fig:Vary_tip_sample_Moore}(b) we instead used eq.~\ref{eq:DeltaMz-strong-B1}.

The Fig.~\ref{fig:Vary_tip_sample_Moore}(a) calculation yields a much lower average magnetization and, therefore, predicts a larger signal. 
To get the Fig.~\ref{fig:Vary_tip_sample_Moore}(a) calculation to agree with the measured spin signal, it was necessary to assume either a $B_1$ far below that expected based on the input power or a much stiffer spring constant than that inferred from analysis of thermomechanical fluctuations. 
In the Fig.~\ref{fig:Vary_tip_sample_Moore}(a) calculation we used a spring constant of $k_c = \SI{3.9}{\milli\newton\per\meter}$, five times the measured value. 
No choice of $B_1$ and $k_c$ gave a calculated signal matching both the bulk and local peaks simultaneously, and as the tip-sample separation changed, the inferred $B_1$ and $k_c$ would have to change as well to maintain good agreement with the measured spin signal.

Calculating the signal using the Sec.~\ref{sec:intermittent-strong} formulas resolved these discrepancies.
The  Fig.~\ref{fig:Vary_tip_sample_Moore}(b) signal calculations used the measured $k_c = \SI{0.8}{\milli\newton\per\meter}$ and our best estimate of $B_1 = \SI{2.9}{\micro\tesla}$ at the 10 dBm (10 mW) input power. 
These simulations achieved quantitative agreement with experiment with independently calibrated parameters and no per-curve adjustment.
In Fig.~\ref{fig:Vary_tip_sample_Moore}(b), we can see that the local peak is comparatively smaller than the bulk peak at small tip-sample separation $h$.
This is because the local spins are saturated less effectively, due to tip motion, in the larger tip-field
gradient present at small $h$.

\begin{figure}
  \centering
  \includegraphics[width=3.25in]{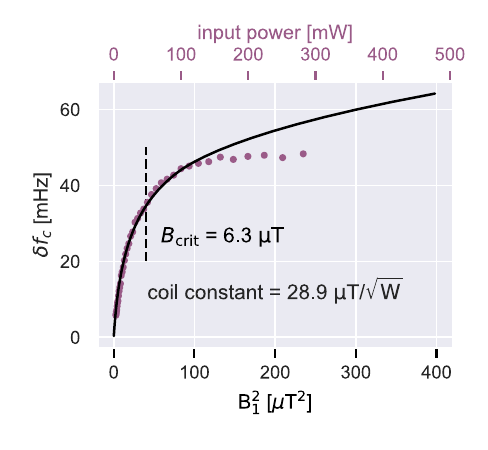}
    \vspace{-2em}
    \caption{Calculated (solid line) and measured (circles) spin signal \latin{vs}.\ microwave power
    at $B_0 = \SI{642}{\milli\tesla}$ and 
    $h = \SI{150}{\nano\meter}$.
      The upper $x$-axis is the estimated total power into the microwave resonator.
      The lower $x$-axis is derived from the $B_1$ used in the simulation. 
      A coil constant of \SI{28.9}{\micro\tesla \: \watt^{-1/2}} was inferred by requiring the calculated signal plotted versus both the upper and lower $x$ axes to agree.  
      This coil constant was used for subsequent simulations. 
      We note that the $B_1 \ll B_1\supr{crit}$ approximation may no longer hold beyond $B_1 = \SI{6.3}{\micro\tesla}$ (dashed line).                  
    }
\label{fig:Vary_power_fit_cP}
\end{figure}

The coil constant of the microwave resonator was obtained by analyzing the spin signal \latin{vs}.\ microwave power data, Fig.~\ref{fig:Vary_power_fit_cP}.
Microwaves excited the cantilever directly, producing a spurious signal at $f\subs{mod}$. 
An off-resonance control signal was therefore measured at $B_0 = \SI{550}{\milli\tesla}$ and subtracted from the measured bulk resonance peak at $B_0 = \SI{640}{\milli\tesla}$ (for $h = \SI{150}{\nano\meter}$). 
The simulation predicts that the spin signal should increase continuously with increasing microwave power due to power broadening. 
In the experiment, however, the observed spin signal levels off above \SI{100}{\milli\watt}. 
In our analysis of saturation in the presence of a time-dependent offset \cite{Boucher2023sep}, the rate at which magnetization saturates was revealed to be $r = 1 \big/ 2 T_2$ in the infinite-$B_1$ limit. 
We estimate that this limited saturation rate is an unlikely explanation for the leveling-off of the spin signal. 
A more plausible explanation is that the leveling-off was the result of sample heating at high power, which decreases the signal by reducing the equilibrium Curie-law magnetization. 

For an untuned transmission line with the same dimensions as the resonator, finite element modeling using the ANSYS package predicts a coil constant \SI{36.8}{\micro\tesla \: \watt^{-1/2}}; see Ref.~\citenum{Moore2011sep}, Sec.~2.4.2.
The Fig.~\ref{fig:Vary_power_fit_cP} fit gives a coil constant of $c_P = \SI{28.9}{\micro\tesla \: \watt^{-1/2}}$, in comparatively reasonable agreement with the ANSYS model.
In contrast, calculating signal using the steady-state approximation, as in Ref.~\citenum{Moore2009dec},  gives $c_P = \SI{1.4}{\micro\tesla \: \watt^{-1/2}}$, in comparatively poor agreement with the ANSYS model.
The microwave resonator had a circuit quality factor, measured outside the magnetic resonance force microscope, of $Q \sim 800$; no electrical resonance was observed with the resonator in the microscope.  
We hypothesize that this disappearance of the electrical resonance was due to either a poorly defined electrical ground in the microscope or the sample's metal coating, damping the electrical quality factor.
Assuming a well-matched resonator, we would expect the measured coil constant to be $\sqrt{Q} = 28$ larger than that predicted by the ANSYS model.
This enhancement was not observed; the measured coil constant is consistent with $Q \sim 1$.

\begin{figure}
    \centering
    \includegraphics[width=3.25in]{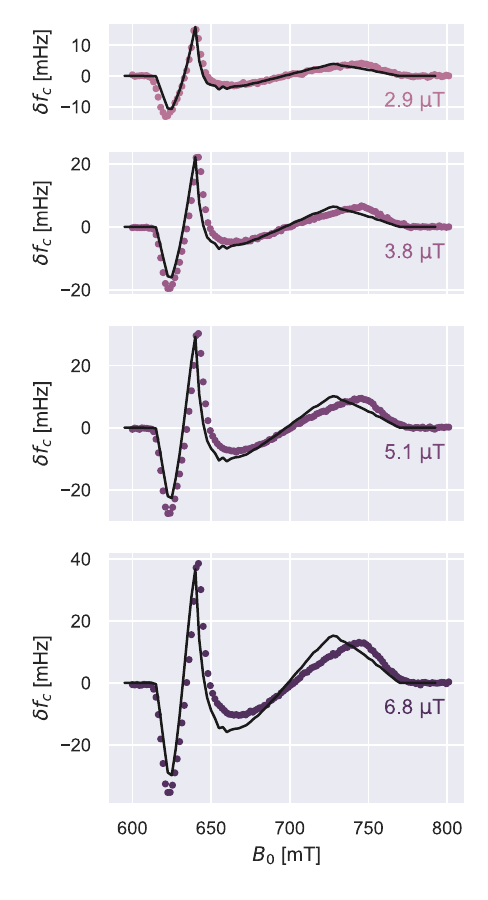}
    \vspace{-1.5em}
    \caption{Measured (circles) and calculated (lines) spin signal \emph{vs}.\ magnetic field at four microwave powers (circles): 10 mW, 18 mW, 32 mW, and 56 mW of microwave power accounting for the amplifier and cabling. The simulation $B_1$ values are displayed on the plot.
    Experimental parameters: $h = \SI{150}{\nano\meter}$, $k_c = \SI{0.8}{\milli\newton\per\meter}$. 
}
\label{fig:Vary_power_vs_field}
\end{figure}

Spin signal \latin{vs.}\ magnetic field data were collected at selected values of microwave power, Fig.~\ref{fig:Vary_power_vs_field}. 
Calculations capture the dependence of signal size and lineshape on power, although at high power the calculated local-peak signal (at $B_0 = 650$ to $\SI{750}{\milli\tesla}$) diverges somewhat from the measured signal.
We note that this divergence is not due to heating, which would affect the bulk and local peaks equally.

\begin{figure*}
  \centering
  \includegraphics[width=6.5in]{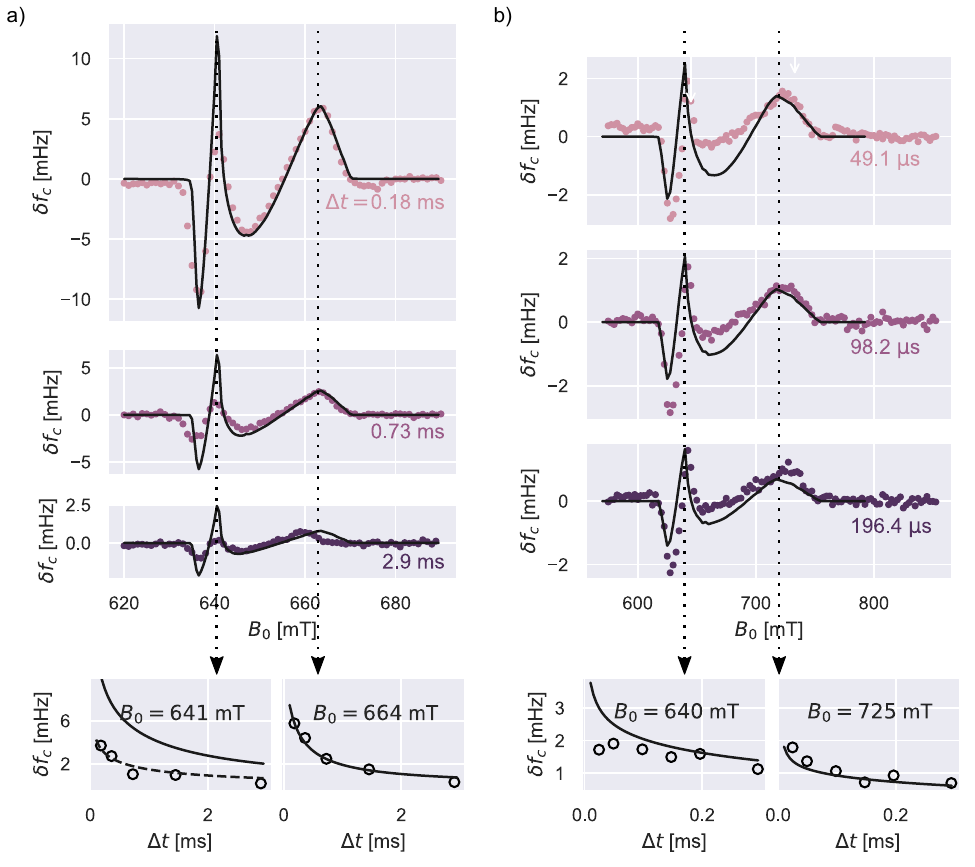}
  \vspace{-1em}
  \caption{Measured (circles) and calculated (lines) spin signal \emph{vs}.\ magnetic field for indicated values of the microwave burst spacing $\Delta t = n T\subs{c}$.
    Experimental conditions: 
    (a) The cantilever was driven at its first mechanical resonance, $f_c = \SI{5.512}{\kilo\hertz}$, with $h = \SI{1027}{\nano\meter}$, $B_1 = \SI{2.9}{\micro\tesla}$, and $n = 1$, $4$, and $16$.
    (b) The cantilever was driven at its second mechanical resonance, $f_c = \SI{40.727}{\kilo\hertz}$, $k_c = \SI{100}{\milli\newton\per\meter}$, with $h = \SI{150}{\nano\meter}$, $B_1 = \SI{11.5}{\micro\tesla}$, and $n = 2$, $4$, and $8$.
    Bottom: Measured (circles) and calculated (line) signal vs.\ pulse spacing at two $B_0$ values.
    The dashed line in (a) is the calculated signal divided by three, to aid in comparing the reduction in signal with longer spacing between pulses.
  }
\label{fig:Vary_interpulse_spacing}
\end{figure*}

Spin signal \latin{vs.}\ magnetic field data were collected as a function of the spacing $\Delta t = n T\subs{c}$  between microwave pulses, Fig.~\ref{fig:Vary_interpulse_spacing}, with the cantilever driven at its first and second mechanical resonance.
For the first cantilever mode, signal was collected at $h = \SI{1027}{\nano\meter}$, Fig.~\ref{fig:Vary_interpulse_spacing}(a).
As in Fig.~\ref{fig:Vary_tip_sample_Moore}, the measured bulk peak in Fig.~\ref{fig:Vary_interpulse_spacing}(a) is somewhat shorter and broader than the simulation predicts.
The simulations assumed $T_1 = \SI{1.3}{\milli\second}$, the spin-lattice relaxation time inferred from both inductively detected electron spin-echo experiments and cantilever phase-shift MRFM measurements \cite{Moore2009dec}.
As $\Delta t$ is increased in the Fig.~\ref{fig:Vary_interpulse_spacing}(a) experiment, calculations captured the observed reduction of the local-peak signal quite well, even out to the longest delay, $\Delta t = \SI{2.9}{\milli\second} \geq 2 T_1$.
Surprisingly, calculations failed to capture the observed dependence of the bulk-peak signal on time delay $\Delta t$.
We hypothesize that the discrepancy between the height of the bulk peak and the decay of the signal size compared to the simulated values is due to spin diffusion, which would become more of a factor over a long $\Delta t$ and would occur more efficiently in the comparatively smaller magnetic field gradient experienced by bulk spins \cite{Budakian2004jan,Eberhardt2007nov,Vinante2011dec,Isaac2016apr,Isaac2017jun}. A number of the relevant diffusion parameters for our system are collected in Table 2 of ref.~\citenum{Isaac2016apr}. We expect a \SI{275}{\nano\meter} diffusion length in the $T_1$ of our sample electron spins. This is a significant fraction of the \SI{1376}{\nano\meter} magnet radius.  Table 2 also calculates a first critical gradient, $G_1\supr{crit} = \SI{0.6}{\micro\tesla\per\nano\meter}$, above which spin diffusion is reduced, and a second gradient, $G_2\supr{crit} = \SI{50}{\micro\tesla\per\nano\meter}$, above which diffusion is completely suppressed. Using the high-$B_0$ end of the MRFM lineshape in Fig.~\ref{fig:Vary_tip_sample_Moore} we estimate $B_{zz} \approx \SI{-40}{\micro\tesla\per\nano\meter}$ at $h = \SI{1027}{\nano\meter}$ for local spins, enough to almost fully suppress diffusion. Using the low-$B_0$ end of the same MRFM lineshape, we obtain a bulk spin $B_{zz} \approx \SI{8}{\micro\tesla\per\nano\meter}$, enough to reduce but not fully suppress diffusion. To fully understand the effect of spin diffusion on the MRFM signal, however, would require a numerical simulation.

Fig.~\ref{fig:Vary_interpulse_spacing}(b) shows the detection carried out using the cantilever's second mechanical resonance at $f_c = \SI{40.727}{\kilo\hertz}$. Because the cantilever tip moves faster at this higher frequency, the input power to the resonator was increased to $B_1 = \SI{11.5}{\micro\tesla}$ to ensure spin saturation. Spacing between microwave pulses is much shorter at the same multiples of the cantilever period used in (a), but the signal still matches the local peak quite well. The decay of the bulk peak is predicted less accurately than the local peak, possibly because of overlap with the low $B_0$ end of the lineshape. We do not expect as much spin diffusion in Fig.~\ref{fig:Vary_interpulse_spacing}(b) because of the smaller tip-sample separation and shorter delay between microwave pulses.

\begin{figure*}
  \centering
  \includegraphics[width=6.5in]{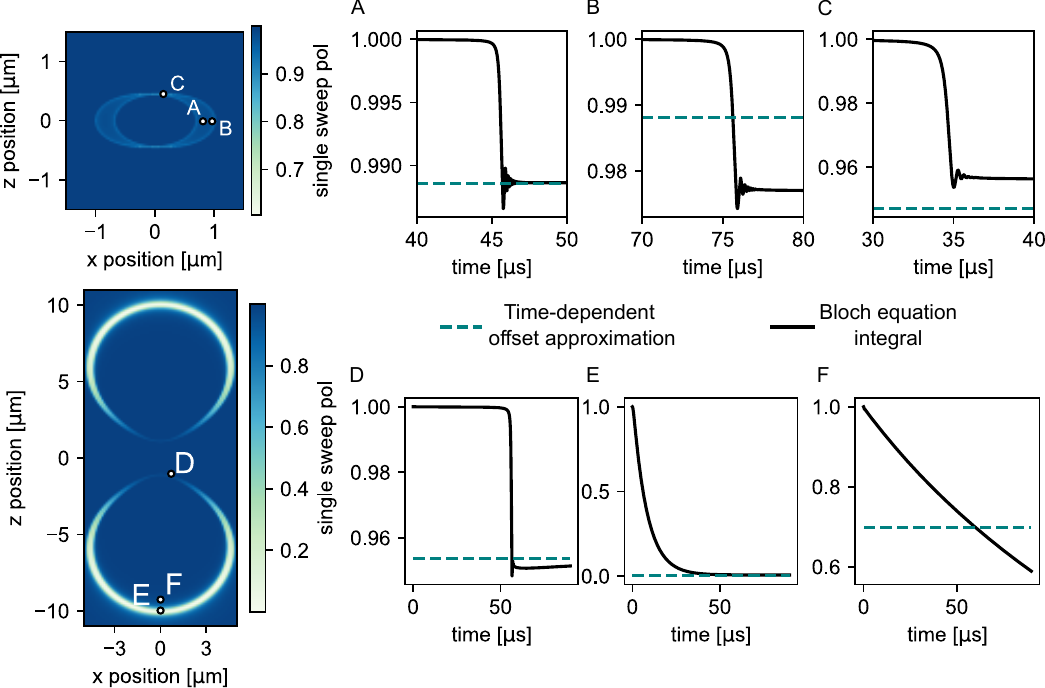}
  \caption{Evaluating the validity of the Sec.~\ref{sec:intermittent-strong} equations.
    Left: Cross-section of the fraction of the initial magnetization remaining after one microwave burst.
    Top row: local peak, $B_0 = \SI{732}{\milli\tesla}$. 
    Bottom row: bulk peak, $B_0 = \SI{645}{\milli\tesla}$. 
    Right: Normalized spin magnetization versus time for spins at three representative locations,  computed by numerically evolving the Bloch equations with a \SI{10}{\pico\second} time step (black solid line) and using low-$B_1$ equations in Sec.~\ref{sec:intermittent-strong} (green dotted line).
    Simulation parameters: uniformly magnetized sphere with 
    radius $r = \SI{1376}{\nano\meter}$ and 
    saturation magnetization $\mu_0 M_\mathrm{s} = \SI{523}{\milli\tesla}$, 
    cantilever zero-to-peak oscillation amplitude $\SI{164}{\nano\meter}$, $h = \SI{150}{\nano\meter}$,
     and $B_1 =  \SI{2.9}{\micro\tesla}$.
     The microwave burst time was \SI{90}{\micro\second}, corresponding to a half-cycle burst.
}
\label{fig:Linearity_plot}
\end{figure*}

The treatment of Sec.~\ref{sec:intermittent-strong} assumed that spins experience a magnetic field changing linearly in time \cite{Boucher2023sep}.
In reality, spins beneath the tip can experience a field varying non-linearly in time.
To evaluate the validity of the Sec.~\ref{sec:intermittent-strong} equations in this situation, we compared the post-burst magnetization predicted by the Sec.~\ref{sec:intermittent-strong} equations to magnetization computed by numerically evolving the Bloch equations with a short time step of  \SI{10}{\pico\second}. 
Results are shown in Fig.~\ref{fig:Linearity_plot}. The left-hand column of Figure~\ref{fig:Linearity_plot} shows the polarization of spins in the top layer of the sample after the first microwave pulse. 
The top row of the figure shows spins that are near the cantilever magnet and contribute to the ``local peak'' at $B_0 = \SI{732}{\milli\tesla}$.
The bottom row shows ``bulk peak'' spins, in resonance at $B_0 = \SI{645}{\milli\tesla}$,  which experience a much weaker tip field. 
Spins A and B experience a tip field that changes linearly with cantilever $x$ position.

What do we expect to see?
The cantilever tip motion is sinusoidal, therefore slowing down at the apex.
Spins B and C that pass through resonance near the apex of cantilever motion will remain in resonance longer and be better saturated than spins that pass through resonance when the tip is at its maximum velocity, such as A.
These spins are far fewer in number than those spins that experience a linearly changing tip field. 
Spins at the bottom or sides of the resonant slice (C, D, and F) will also experience a tip field that varies non-linearly in time, due to the curvature of the tip magnetic field there. 
However, the resonant slice is narrow relative to the distance of tip motion, so spins at the bottom or sides of the slice are likewise fewer in number than other types of active spins. 
Spins E and F experience a very slow change in tip field with time, and spend the entire microwave pulse near resonance.
While these spins do not experience a tip field varying linearly in time, the Sec.~\ref{sec:intermittent-strong} equations nevertheless correctly predict that spins E and F will become almost completely saturated.
So while we expect the Sec.~\ref{sec:intermittent-strong} equations to poorly describe some spins in the sample, these spins make up a small minority that do not contribute much to the signal.

The code developed for the Fig.~\ref{fig:Vary_tip_sample_Moore} through \ref{fig:Linearity_plot} large-tip calculations can now be applied to assess signal loss in small-tip experiments.
In these experiments, we hypothesize additional signal loss from tip-related spin relaxation.
Signal loss from $T_1$ reduction can be modeled using the equations presented in Sec.~\ref{sec:intermittent-strong-short}.
Searching for signal from short-$T_1$ electrons, and modeling this signal with the Sec.~\ref{sec:intermittent-strong-short} equations, is beyond the scope of this paper.
Nevertheless, in the following section, we show that the Sec.~\ref{sec:intermittent-strong-short} equations can help us understand and eliminate spurious signals.

\subsection{Spurious signal}

Spurious signals are a challenge in MRFM experiments.
Applied radiowave (RF) or microwave (MW) irradiation can create time-dependent forces and force gradients that mimic spin signal.
These spurious signals are present whether or not the spins are in resonance with the applied magnetic field. 

A number of strategies have been invented to mitigate these spurious signals.
In measurements of Curie-law nuclear spin magnetization, Degen found that a spurious RF-induced force signal could be mitigated by ``preheating'' the cantilever, applying an off-resonant RF pulse before the resonant spin excitation pulse \cite{Degen2005}. 
Similarly, Krass and coworkers avoided spurious excitation of the cantilever by applying off-resonant pulses to maintain constant RF power between spin inversion sequences \cite{Krass2024oct}. 
In their measurement of nuclear spin noise, Longenecker \emph{et al}.\ observed spurious cantilever excitation from the triangle-wave, frequency-sweep RF irradiation applied twice per cantilever cycle in their experiment; they eliminated this excitation by introducing a compensating amplitude modulation of the RF amplitude \cite{Longenecker2012nov}.

\begin{figure*}
    \centering
    \includegraphics[width=5.50in]{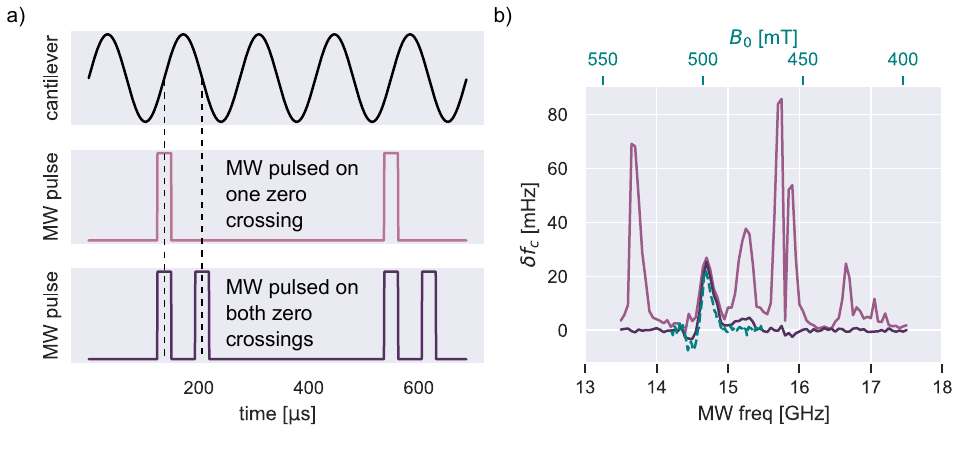}
    \caption{Observation and mitigation of a spurious frequency-shift signal in a force-gradient magnetic resonance force microscope experiment.
    (a) Microwave timing diagram.
    Microwave bursts were applied every third cycle, 
    at either one zero-crossing per cycle (middle, magenta line)
    or at two zero-crossings per cycle (lower, purple line). 
    (b) Cantilever frequency shift \latin{vs.}\ 
    microwave frequency (lower $x$-axis) 
    at $B_0 = \SI{516}{\milli\tesla}$,
    for microwaves applied at one zero crossing (solid magenta line) 
    or at both zero crossings (solid purple line).
    Cantilever frequency shift \latin{vs.}\
    \latin{vs.}\ magnetic field (upper $x$-axis) 
    at $f\subs{MW} = \SI{14.2}{\giga\hertz}$ (dashed green line). 
    Experimental parameters: 
    cantilever period $T\subs{c} = \SI{137}{\micro\second}$, 
    microwave burst length \SI{25}{\micro\second},  
    $h = \SI{50.2}{\nano\meter}$, 
    $B_1 = \SI{40}{\micro\tesla}$, and
    $x_{\mathrm{0pk}} = \SI{164.0}{\nano\meter}$. The sample was a 300 nm thick polystyrene film doped to 40 mM with 4-amino-TEMPO.   
}
\label{fig:One_vs_both_crossings}
\end{figure*}

A spurious signal is observed in modulated-CERMIT electron spin resonance measurements as well, Fig.~\ref{fig:One_vs_both_crossings}.
The spin resonance peak is revealed by fixing the microwave frequency at $f\subs{MW} = \SI{14.2}{\giga\hertz}$ and scanning magnetic field, Fig.~\ref{fig:One_vs_both_crossings}(b, dashed green line); a spin resonance peak is apparent near $B_0 = \SI{500}{\milli\tesla}$ in the figure.
Field-induced motion of the scanning probe microscope probe head often makes scanning the magnetic field problematic, however.
In this situation, it is preferable to scan the microwave frequency.
Frustratingly, applying microwave bursts at a cantilever zero crossing every three cantilever cycles (to minimize heating) gave a cantilever frequency shift \latin{vs.}\ microwave frequency signal with many false peaks, Fig.~\ref{fig:One_vs_both_crossings}(b, solid magenta line).
We hypothesize that the false peaks correspond to resonances in the experiment's imperfectly matched coplanar waveguide.

\begin{figure*}
  \centering
  \includegraphics[width=5.00in]{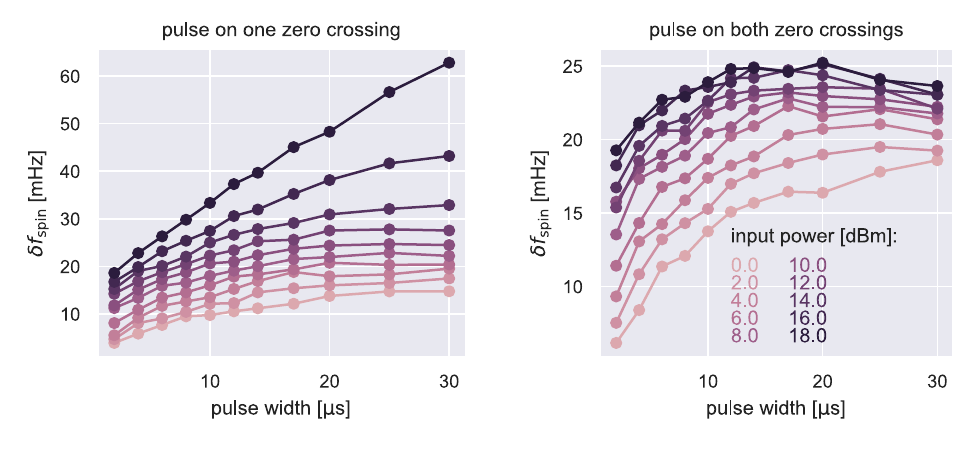}
  \caption{Spin signal \latin{vs.}\ microwave burst duration (i.e., pulse width), for microwaves applied once (left) and twice (right) per cycle.
    The microwave power is given in the legend.
    Experimental parameters:
    $B_0 = \SI{516}{\milli\tesla}$, 
    $h = \SI{50.2}{\nano\meter}$, 
    $x\subs{0-pk} = \SI{164.0}{\nano\meter}$, 
    and $f\subs{MW} = \SI{14.7}{\giga\hertz}$. The sample was a 300 nm thick polystyrene film doped to 40 mM with 4-amino-TEMPO.               
  }
\label{fig:Pulse_width_and_power_vs_lockin_signal}
\end{figure*}

Remarkably, the false frequency-shift peaks are eliminated by applying microwave bursts at alternating zero crossings, Fig.~\ref{fig:One_vs_both_crossings}(b, solid purple line).
In Fig.~\ref{fig:Pulse_width_and_power_vs_lockin_signal} we plot the spin-induced frequency shift \latin{vs.}\ the duration of the microwave burst.
Applying bursts at one zero crossing gave a signal which increased monotonically with burst duration, Fig.~\ref{fig:Pulse_width_and_power_vs_lockin_signal}(a).
Applying bursts on alternate zero crossings, on the other hand, gave a signal showing the expected saturation at long burst duration, Fig.~\ref{fig:Pulse_width_and_power_vs_lockin_signal}(b).
The monotonic increase of signal in Fig.~\ref{fig:Pulse_width_and_power_vs_lockin_signal}(a) is presumably due to a spurious excitation of the cantilever by the applied microwaves.

The equations of Sec.~\ref{sec:intermittent-strong-short} explain why applying microwaves at both cantilever zero crossings eliminated the spurious frequency-shift signal.
We hypothesize that microwaves created a spurious force $\Delta F\subs{MW}$ and force gradient $\Delta k\subs{MW}$.
We can use eqs.~\ref{eq:Df_vary_phase} and \ref{eq:Df_half_delay} and the following correspondences to predict how this force and force gradient will affect the cantilever frequency:
\begin{subequations}
\begin{align}
G_1 (M_z\supr{ss} - M_0) & \rightarrow \Delta F\subs{MW} \text{ and}\\
G_2 (M_z\supr{ss} - M_0) & \rightarrow \Delta k\subs{MW}
\label{eq:DF-MW}
\end{align}
\end{subequations}
and the rate $r_1$ representing the lifetime of the spurious microwave-induced force and force gradient. In the case of heating-induced forces and force gradients, for example, the lifetime would depend on the cantilever tip's thermal conductivity.
The resulting modified eq.~\ref{eq:Df_half_delay} predicts that pulsing on alternate zero crossings removes the $\Delta F\subs{MW}$-dependent term from $\Delta f\subs{spin}$, consistent with the Fig.~\ref{fig:One_vs_both_crossings} experiment finding that applying a microwave at two zero crossings eliminates the spurious frequency-shift signal.
The large difference between the solid-magenta and solid-purple signals in Fig.~\ref{fig:One_vs_both_crossings}(b) indicates that $\Delta F\subs{MW}$ is large, while the similarity in the solid-purple and dotted-green signals in Fig.~\ref{fig:One_vs_both_crossings}(b) indicates that $\Delta k\subs{MW}$ is negligibly small.

\begin{figure}
  \centering
  \includegraphics[width=3.50in]{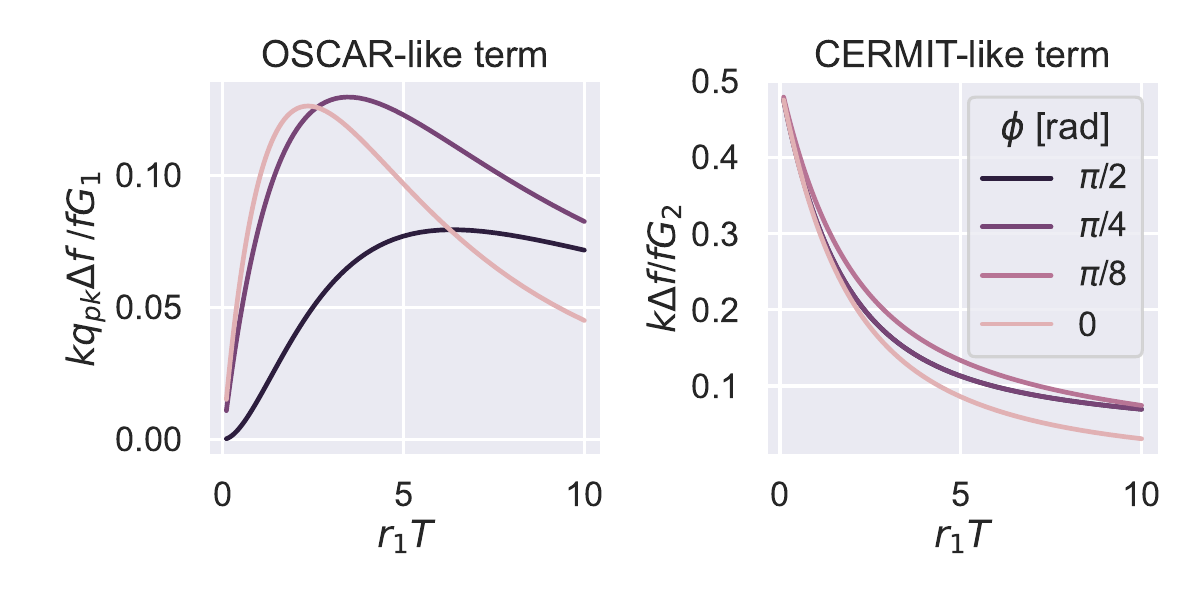}
  \caption{Magnitude of the (first) OSCAR-like and the 
  (second) CERMIT-like terms in eq.~\ref{eq:Df_vary_phase}, with $n = 1$, as a function of phase, $\phi$, and $T_1$-normalized time delay between microwave bursts, $r_1 T$.
  A phase delay of $\phi = 0$ and $\phi = \pi/2$ corresponds to applying microwaves at the cantilever zero crossing and apex, respectively.
  }
\label{fig:phase_dependent_decay}
\end{figure}

Equation~\ref{eq:Df_vary_phase} predicts how the spin signal changes for short-$T_1$ electrons with increasing time delay $T$ between microwave pulses (as in the Fig.~\ref{fig:Vary_interpulse_spacing} experiment). 
For $T_1 < T$, both OSCAR-like and CERMIT-like terms contribute to the frequency shift in a phase-dependent way. 
In Fig.~\ref{fig:phase_dependent_decay} we plot the magnitude of the two terms as a function of the $T_1$-normalized time delay between microwave bursts, $r_1 T$.
As the burst delay $T$ increases (or equivalently, $T_1$ decreases), the force-based OSCAR-like term increases, reaches a maximum, then decreases; in contrast, the force-gradient-based CERMIT-like term simply decreases.
This behavior is phase-dependent in both cases, but the phase dependence is more pronounced for the OSCAR-like term.
In the small-$T_1$ limit, the OSCAR-like term decreases as $\propto T_1$ or $\propto T_1^2$, depending on phase, while the CERMIT-like term decreases primarily as $\propto T_1$.

\section{Conclusions}

In Ref.~\citenum{Boucher2023sep} we tested Eqs.~\ref{eq:alphas}, \ref{eq:lowB1}, and \ref{eq:L-Z-guess} in $^1$H nuclear magnetic resonance experiments on dissolved polystyrene, adjusting the frequency-sweep rate to vary $\alpha_1$ and $\alpha_2$.
Here we showed, Fig.~\ref{fig:Vary_tip_sample_Moore}, that adiabatic losses lead to a signal deficit in magnetic resonance force microscope experiments carried out on electron spins at low temperature with a micron-scale magnetic tip.
We used the Ref.~\citenum{Boucher2023sep} results to write equations for steady-state magnetization that account for spin relaxation between microwave bursts.
New equations for spin signal in the magnetic resonance force microscope experiment, valid in the $T_1 \gg T\subs{c}$ limit, were obtained.

These equations, Secs.~\ref{sec:intermittent-weak} and \ref{sec:intermittent-strong}, correctly predicted the dependence of electron-spin-induced cantilever frequency shift on tip-sample separation, microwave burst delay, and irradiation intensity.
The equations were validated by comparing with magnetization computed by numerically integrating the time-dependent Bloch equations.
The equations were used to compute the signal over a grid of $10^7$ sample points in a few seconds, compared to seconds per grid point for integrating the time-dependent Bloch equations.

Additional equations for spin signal were derived, Sec.~\ref{sec:intermittent-strong-short}, valid in the $T_1 \ll T\subs{c}$ limit.
These equations were used to explain why applying microwaves at alternate zero crossings eliminated a spurious frequency-shift signal arising from microwave-induced forces.
It has been hypothesized that the electron spin signal in small-tip experiments is smaller than expected because of tip-induced spin relaxation \cite{Boucher2023jan}.
The Sec.~\ref{sec:intermittent-strong-short} equations will be useful for designing magnetic resonance force microscope experiments optimized to detect electrons with $T_1$ shortened either by tip-induced fluctuations or by other means \cite{Haan2015dec}. 

Attempts in our laboratory to implement Nguyen's proposal \cite{Nguyen2018feb} with \SI{100}{\nano\meter}-diameter cobalt tips have so far been unsuccessful, with the observed electron spin signals up to 20-fold smaller than expected \cite{Boucher2023jan}. Work here shows that part of this signal deficit can be attributed to the swept-field breakdown of saturation.

\section*{Acknowledgments}

Research reported in this publication was supported by Cornell University, the Army Research Office under Award Number W911NF-17-1-0247, and the National Institute of General Medical Sciences of the National Institutes of Health under Award Number R01GM143556.
The content of this manuscript is solely the responsibility of the authors and does not necessarily represent the official views of Cornell University, the U.S.\ Army Research Office, the National Institutes of Health, the Department of the Army, the Department of Defense, or the U.S.\ Government.
The authors have no competing interests to declare.

\bibliographystyle{bst/elsarticle-num-names} 

\end{document}